\documentclass[11pt]{article}
\usepackage[dvipdfmx]{graphicx}
\usepackage{bm}
\usepackage{amsmath}
\usepackage{amssymb}
\usepackage{float}
\usepackage{url}
\newcommand{\nn}{\nonumber\\}
\newcommand{\bra}[1]{\left<#1 \right|}
\newcommand{\ket}[1]{\left| #1 \right>}
\renewcommand{\thepage}{}
\makeatletter
\@addtoreset{equation}{section}
\renewcommand{\theequation}{\thesection.\@arabic\c@equation}
\makeatother
\renewcommand{\thefootnote}{\fnsymbol{footnote}}
\begin{document}
\begin{titlepage}
\title{
\vspace*{-4ex}
\hfill{\normalsize UTHEP-717}\\
\hfill
\begin{minipage}{3.5cm}
\end{minipage}\\
\bf Vector profile and gauge invariant observables of string field
 theory solutions for constant 
 magnetic field background
\vspace{0.5em}
}

\author{
Nobuyuki {\sc Ishibashi}$^{1}$\footnote{ishibash@het.ph.tsukuba.ac.jp},
\ \ Isao {\sc Kishimoto}$^{2}$\footnote{ikishimo@ed.niigata-u.ac.jp},
\ \ Toru {\sc Masuda}$^{3,4}$\footnote{masudatoru@gmail.com}
\ \ and\ \  
Tomohiko {\sc Takahashi}$^{5}$\footnote{tomo@asuka.phys.nara-wu.ac.jp}
\\
\vspace{0.5ex}\\
$^{1}${\it Tomonaga Center for the History of the Universe,}\\ 
{\it University of Tsukuba}\\
{\it Tsukuba, Ibaraki 305-8571, Japan}
\vspace{1ex}
\\
$^{2}${\it Faculty of Education, Niigata University,}\\
{\it Niigata 950-2181, Japan}
\vspace{1ex}
\\
$^{3}${\it CEICO, Institute of Physics of the Czech Academy of
 Sciences,}\\
{\it Na Slovance 1999/2, 182 21  Prague 8, Czech Republic}
\vspace{1ex}
\\
$^{4}${\it CORE of STEM, Nara Women's University,}\\
{\it Nara 630-8506, Japan}
\vspace{1ex}
\\
$^{5}${\it Department of Physics, Nara Women's University,}\\
{\it Nara 630-8506, Japan}
\vspace{2ex}
}

\date{}
\maketitle

%
\vspace{7ex}

\begin{abstract}
\normalsize
We study profiles and gauge invariant observables of
classical solutions corresponding to a constant magnetic field on
a torus in open string field theory. We numerically find that the
profile is not discontinuous on the torus, although the solution describes
topologically nontrivial configurations in the context of low energy
effective theory. From the gauge invariant observables, we show that
the solution provide correct couplings of closed strings to a D-brane
with constant magnetic field.
\end{abstract}

\vspace{5ex}

\end{titlepage}

\renewcommand{\thepage}{\arabic{page}}
\renewcommand{\thefootnote}{\arabic{footnote}}
\setcounter{page}{1}
\setcounter{footnote}{0}
%
\tableofcontents

\section{Introduction}
String field theory is believed to admit classical solutions that
describe a wide range of moduli space of string theory. In bosonic
open string field theory, Erler and Maccaferri\cite{Erler:2014eqa}
proposed a way to construct classical solutions representing any
time-independent
open string background by use of boundary condition changing (BCC)
operators. Following their method, a solution corresponding to constant
magnetic field background has been constructed by some of the present
authors in~\cite{Ishibashi:2016xak}. It was found that the classical
action of the solution calculated from the operator product expansions
(OPEs) of BCC operators agrees with the Dirac-Born-Infeld action.

This magnetic solution has several new features compared with the
solutions discovered so far. Firstly it has no direct relation to
tachyon dynamics, such as tachyon condensation or tachyon lump, or
it is not obtained simply by marginal deformation of already-known
solutions. Secondly, the solution on a torus corresponds to the configuration
has a nonvanishing Chern number of the $U(1)$ gauge field. Such a
topologically nontrivial solution had never been constructed before
in string field theory.

The purpose of this paper is to investigate some physical properties
of the magnetic solution. Firstly, we will study the so-called tachyon
and vector profiles of the solution defined as follows. Using the
Fock space expression, a solution $\ket{\Psi}$ of the open string
field theory can be expanded as 
\begin{eqnarray}
 \ket{\Psi}&=& \sum_{\bm p}t_{\bm p}\ket{T_{\bm p}}+
\sum_{\bm p}{A_{\bm p}}^\mu \big|{V_{\bm p}}_\mu\big>+\cdots,
\label{eq:Fockrep}
\end{eqnarray}
where the lower mass states, corresponding to the tachyon and the
massless vector field are expressed as 
\begin{eqnarray}
&&
 \ket{T_{\bm p}}= c_1\ket{\bm p}
 = ce^{i{\bm p}\cdot {\bm X}}(0)\ket{0},
\\
&&
 \big|{V_{\bm p}}^\mu\big>
= c_1\alpha_{-1}^\mu\ket{\bm p}
= i\sqrt{\frac{2}{\alpha'}}\,c\partial
X^\mu e^{i{\bm p}\cdot {\bm X}}(0)\ket{0}.
\end{eqnarray}
The position representation of the component fields is given by the
Fourier transform of $t_{\bm{p}},A_{\bm{p}}^{\mu},\cdots$: 
\begin{eqnarray}
 t({\bm x})=\sum_{\bm p}t_{\bm p}e^{i{\bm p}\cdot {\bm x}},
~~~
A^\mu({\bm x})=\sum_{\bm p}A^\mu_{\bm p}e^{i{\bm p}\cdot {\bm x}},
~~~\cdots.
\label{eq:FourierSum}
\end{eqnarray}
$t({\bm{x}})$ and $A^{\mu}({\bm{x}})$ are called the tachyon and
vector profiles of the solution respectively.

In conventional field theory, we need to divide the torus into some
patches to describe the $U(1)$ gauge field with a constant magnetic
field on a torus, so that the smooth gauge fields on different patches
are related by gauge transformations. Therefore it is natural to ask
whether the magnetic solution needs multiple patches in string
field theory. This question may be examined by evaluating the vector
profile $A^{\mu}({\bm{x}})$ corresponding to the $U(1)$ gauge field.
If $A^{\mu}({\bm{x}})$ has discontinuities, we need to divide the
torus by coordinate patches to represent the configuration by smooth
gauge fields. 

Secondly, we will evaluate gauge invariant observables for the magnetic
solution. In conventional field theory, topologically non-trivial
configurations are characterized by some gauge invariant quantities
which take discrete values. Such quantities have not been found in
string field theory. Instead, we have the gauge invariant observable
which is associated with on-shell closed string vertex operators \cite{Zwiebach:1992bw}:
\begin{eqnarray}
 O_V(\Psi)\equiv \bra{I}c(i)c(-i)V(i,-i)\ket{\Psi},
\label{eq:ginv}
\end{eqnarray}
where $I$ is the identity string field, $V(z,\bar{z})$ denotes an
on-shell closed string vertex operator and $z=+i,\,\bar{z}=-i$ in the complex
plane correspond to the midpoint $\sigma=\pi/2$ of the open string.
If $\Psi$ is a classical solution, the observable represents a coupling
of the closed string vertex operator $V$ to the D-brane to which
the solution corresponds. Accordingly, we expect that the observable
for a massless anti-symmetric tensor vertex has a non-trivial value
since the corresponding D-brane has constant background magnetic field.
In order to confirm the existence of background magnetic field and
find a clue for topological invariants in string field theory, we
calculate the gauge invariant observables with massless and zero-momentum
closed string vertex operators.

This paper is organized as follows. In section 2, we briefly review
the constant magnetic solution on torus. In section 3, we study tachyon
and vector profiles of the solution. As a by-product, we show periodic
and quasi-periodic properties of the solutions. In section 4, we evaluate
the gauge invariant observables of the solution. By comparing the
resulting observables with the Dirac-Born-Infeld (DBI) action, we find
that the solution indeed corresponds to constant magnetic field background.
In section 5, we will give concluding remarks. In the appendices,
details of calculations are exhibited.

\section{Classical solutions for constant magnetic field background}

We would like to consider the configuration with a constant background
$F_{\mu\nu}$. We concentrate on the spatial directions $X^{1}$ and
$X^{2}$, since a real antisymmetric tensor $F_{\mu\nu}$ can be transformed
into a block diagonal form with $2\times2$ blocks. Let us consider
the bosonic open string field theory in which these spatial directions
are toroidally compactified with radii $R_{1}$ and $R_{2}$, and
the Neumann boundary conditions are imposed on the variables $X^{1},X^{2}$.
The time direction $X^{0}$ is required to be noncompact in order
to construct the solution following Erler-Maccaferri's method\cite{Erler:2014eqa},
but other directions are unspecified here.

To find the classical solution corresponding to a constant $F_{12}$
background, we need to prepare the BCC operators which changes the
open string boundary conditions for $X^{1},X^{2}$ from the Neumann
boundary condition to the one with $F_{12}$ and vice versa. Such
operators correspond to the open strings with one edge with the free
boundary conditions and the other coupled to the constant magnetic
field. The zero mode coordinates $x^{1},x^{2}$ of these open strings
become noncommutative and we need to introduce the following operators
\[
U=\exp\left(i\frac{x^{1}}{R_{1}}\right),\ \ \ V=\exp\left(i\frac{x^{2}}{R_{2}}\right)\,,
\]
which satisfy 
\[
UV=e^{i\frac{2\pi}{N}}\,VU\,,
\]
where the integer $N$ is related to the magnetic field through the
Dirac quantization condition: 
\begin{eqnarray}
 (2\pi)^2R_1R_2F_{12}=2\pi N.
  \label{Dirac}
\end{eqnarray}
The zero mode algebra has a $|N|$ dimensional representation. Correspondingly,
we can find $\left|N\right|$ pairs of BCC operators: $\sigma_{*}^{k},\ \bar{\sigma}_{*}^{k}\ \ (k=1,\cdots,|N|)$
\cite{Ishibashi:2016xak}, which are primary fields with conformal
weight 
\begin{eqnarray}
 h=\frac{\lambda(1-\lambda)}{2},~~~(\ \tan\pi\lambda=2\pi\alpha'F_{12}\ ).
\end{eqnarray}
Following Erler-Maccaferri's method, we multiply $\sigma_{*}^{k},\ \bar{\sigma}_{*}^{k}$
by the vertex operators $e^{\pm i\sqrt{h}X^{0}}$ and appropriate
normalization factors and construct $|N|$ pairs of modified BCC operators
$\sigma^{k},\ \bar{\sigma}^{k}$. They are primary fields with conformal
weight zero and satisfy the OPEs 
\begin{eqnarray}
 \bar{\sigma}^k(s)\sigma^l(0)\sim \delta_{k,l},~~~
 \sigma^l(s)\bar{\sigma}^k(0)\sim \frac{\delta_{l,k}}{|\cos\pi\lambda|}
=\frac{\delta_{l,k}}{\sqrt{1+(2\pi\alpha'F_{12})^2}},
\label{eq:OPEsigma}
\end{eqnarray}
for small positive $s$.

Having these BCC operators, the classical solution corresponding to
the constant magnetic field background \cite{Ishibashi:2016xak} are
given as follows: 
\begin{eqnarray}
 \Psi_0 &=& \Psi_{\rm tv}+\sum_{k,l}A_{k,l}\Phi^{k,l},
\label{eq:sol}
\end{eqnarray}
where $\Psi_{\rm tv}$ is the Erler-Schnabl solution for the tachyon
vacuum \cite{Erler:2009uj}, 
\begin{eqnarray}
  \Phi^{k,l} &=& -\frac{1}{\sqrt{1+K}}c(1+K)\sigma^k
\frac{B}{1+K}\bar{\sigma}^l(1+K)c\frac{1}{\sqrt{1+K}},
\end{eqnarray}
and $A_{k,l}$ is a hermitian $|N|\times |N|$ matrix
satisfying $A^2=A$. The second term is a solution to the equations of
motion around the tachyon vacuum. 
Suppose that  $A$ is given by
\begin{eqnarray}
 A={\rm diag}(\,\underbrace{1,\,\cdots,\,1}_M,\,
\underbrace{0,\,\cdots,\,0}_{|N|-M}\,),
\label{eq:Adiag}
\end{eqnarray}
then the solution can be regarded as describing $M$ D-branes with
magnetic field condensation. Using the OPEs (\ref{eq:OPEsigma}),
it is easy to show that the energy of the solution is given by $M$
times that of a single D-brane with magnetic field condensation.

For later discussion, it is convenient to rewrite the solution (\ref{eq:sol})
in the following way. By using the algebra of $K$, $B$, $c$ and
$\sigma^{k}$, $\bar{\sigma}^{l}$, the solution $\Phi^{k,l}$ can
be decomposed to three parts: 
\begin{eqnarray}
 \Phi^{k,l}&=&-\frac{\delta_{k,l}}{|\cos\pi\lambda|}\Psi_{\rm tv}
+\phi_1^{k,l}+Q_{\rm B}\phi_2^{k,l},
\label{eq:Phi}
\end{eqnarray}
where $\phi_1^{k,l}$ and $\phi_2^{k,l}$ are defined as
\begin{eqnarray}
 \phi_1^{k,l}&=& -\frac{1}{2}\frac{1}{\sqrt{1+K}}c\partial \sigma^k
\frac{1}{1+K}\bar{\sigma}^l\frac{1}{\sqrt{1+K}}
+\frac{1}{2}\frac{1}{\sqrt{1+K}}\sigma^k
\frac{1}{1+K}\partial \bar{\sigma}^lc\frac{1}{\sqrt{1+K}},
\\
\phi_2^{k,l}&=&
\frac{1}{2}\frac{1}{\sqrt{1+K}}c\partial \sigma^k
\frac{B}{1+K}\bar{\sigma}^l\frac{1}{\sqrt{1+K}}
+\frac{1}{2}\frac{1}{\sqrt{1+K}}\sigma^k
\frac{B}{1+K}\partial \bar{\sigma}^lc\frac{1}{\sqrt{1+K}}.
\end{eqnarray}
It should be noted that $\phi_1^{k,l}$ and $\phi_2^{k,l}$
independently satisfy the string field
reality condition,\footnote{We follow the definition of the conjugate
$\ddagger$ given in \cite{Erler:2014eqa}. $K$, $B$ and $c$ are self-conjugate
and $\sigma^\ddagger=\bar{\sigma}$. In addition, we find
$(\partial \sigma)^\ddagger=-\partial\bar{\sigma}$.}
\begin{eqnarray}
 \Big\{\sum_{k,l}A_{k,l}{\phi_1^{k,l}}\Big\}^\ddagger
= \sum_{k,l}A_{k,l}{\phi_1^{k,l}},
\ \ \ 
 \Big\{\sum_{k,l}A_{k,l}Q_{\rm B}{\phi_2^{k,l}}\Big\}^\ddagger
= \sum_{k,l}A_{k,l}Q_{\rm B}{\phi_2^{k,l}}.
\end{eqnarray}
These are useful in checking the correctness of the profile calculation
presented later.

\section{Profiles of the classical solution}

\subsection{Profiles and dual states}

Now let us study the profiles of the magnetic solution $\Psi_{0}$.
In order to extract momentum space profiles from the solution expanded
as (\ref{eq:Fockrep}), we define the states dual to the tachyon and
massless vector states, $\ket{T_{\bm{p}}}$ and $\big|{V_{\bm{p}}}_{\mu}\big>$,
by 
\begin{eqnarray}
 &&
 \big|\tilde{T}_{\bm p}\big>= \frac{1}{(2\pi)^2R_1R_2}
c_0c_1\ket{-\bm p}
 = -\frac{1}{(2\pi)^2R_1R_2}
c\partial c\,e^{-i{\bm p}\cdot {\bm X}}(0)\ket{0},
\\
&&
 \big|{\tilde{V}_{\bm p}}^\mu\big>
= \frac{1}{(2\pi)^2R_1R_2}
c_0 c_1\alpha_{-1}^\mu\ket{-\bm p}
= -\frac{1}{(2\pi)^2R_1R_2}
i\sqrt{\frac{2}{\alpha'}}\,c\partial c\,\partial
X^\mu e^{-i{\bm p}\cdot {\bm X}}(0)\ket{0}.
\label{eq:Vec}
\end{eqnarray}
These are dual to $\ket{T_{\bm{p}}}$ and $\big|{V_{\bm{p}}}_{\mu}\big>$
in the sense that they satisfy 
\begin{eqnarray}
 \big<\tilde{T}_{\bm p},\,T_{\bm q}\big>=\delta_{{\bm p},{\bm q}},
~~~~
 \big<{\tilde{V}_{\bm p}}^\mu,\,{V_{\bm q}}^\nu\big>=
\delta_{{\bm p},{\bm q}}\,\eta^{\mu\nu},
~~~~
 \big<\tilde{T}_{\bm p},\,V_{\bm q}^\mu\big>=
 \big<{\tilde{V}_{\bm p}}^\mu,\,{T_{\bm q}}\big>=0,
\end{eqnarray}
and they are orthogonal to other higher massive states.
The momentum space profile is derived
from the inner product of the dual state and $\Psi_0$:
\begin{eqnarray}
 t_{\bm p}=\big<\tilde{T}_{\bm p},\,\Psi_0\big>,~~~~
 {A_{\bm p}}^\mu= \big<{\tilde{V}_{\bm p}}^\mu,\,\Psi_0\big>.
\label{eq:momprofile}
\end{eqnarray}

Using (\ref{eq:Phi}), we can see that the profiles (\ref{eq:momprofile})
are decomposed to three parts. The calculations are simplified by
using the following identities:
\begin{eqnarray}
&&
 Q_{\rm B}\big|\tilde{T}_{\bm p}\big>=0,
\label{eq:QT}
\\
&&
 Q_{\rm B}\big|{\tilde{V}_{\bm p}}^\mu\big>
=
-\frac{1}{(2\pi)^2R_1R_2}
i\sqrt{\frac{2}{\alpha'}}\,\frac{i}{2}\alpha'p^\mu\,
c\partial c\,\partial^2 c\,
e^{-i{\bm p}\cdot {\bm X}}(0)\ket{0}.
\label{eq:QV}
\end{eqnarray}
Consequently, for example, we only need to deal with $\Psi_{{\rm tv}}$
and $\phi_{1}^{k,l}$ for the calculation of the tachyon profile.

\subsection{Quasi-periodicity of the solution}

Before starting the calculation of the profiles, we would like to
point out that the profiles satisfy quasi-periodic relations. Here
we deal with the solution corresponding to a single D-brane with constant
magnetic field $F_{12}\neq0$, namely $M=1$ in (\ref{eq:Adiag}).
We can construct $|N|$ independent solutions 
\begin{eqnarray}
 \Psi_0^k=\Psi_{\rm tv}+\Phi^{k,k}~~~(k=1,\cdots,|N|).
\label{eq:kthsol}
\end{eqnarray}

To derive the tachyon profile from $\Psi_0^k$,
we have to calculate the inner products $\big<\tilde{T}_{\bm p},\,\Psi_{\rm
tv}\big>$ and $\big<\tilde{T}_{\bm p},\,\phi_1^{k,k}\big>$
as seen from (\ref{eq:QT}).
The former has been known to be a constant\cite{Erler:2009uj}.\footnote{
\begin{eqnarray*}
 \big<\tilde{T}_{\bm p},\,\Psi_{\rm tv}\big>
&=& \int_0^\infty dt\int_0^\infty ds \frac{e^{-s-t}}{4\pi^2\sqrt{ts}}
(s+t+1)^2\left(1+\cos\frac{\pi(t-s)}{s+t+1}\right)=0.509038\cdots.
\end{eqnarray*}}
The latter can be calculated by rewriting in terms of the correlation
function including the tachyon vertex and the BCC operators, which 
has been derived in \cite{Ishibashi:2016xak}:
\begin{eqnarray}
 \Big<e^{-i{\bm p}\cdot {\bm X}}(\xi)
\sigma_*^k(\xi_1)\bar{\sigma}_*^l(\xi_2)\Big>
&=&\frac{(\xi_1-\xi_2)^{\alpha'{\bm p}^2-\lambda(1-\lambda)}}{
(\xi-\xi_1)^{\alpha'{\bm p}^2}(\xi-\xi_2)^{\alpha'{\bm p}^2}}\,
C_{-\bm p}^{l,k},
\label{eq:3pttac}
\\
C_{-\bm p}^{l,k}&=&
\omega^{\frac{n_1n_2}{2}-n_2 l}\,\delta_{k-l,-n_1\,({\rm mod}N)}
\,\delta^{-\frac{\alpha'{\bm p}^2}{2}},
\label{eq:Cp}
\end{eqnarray}
where $\omega$ is an $N$-th root of unity, $\omega=\exp(2\pi i/N)$,
and $n_{i}$ are momentum quantum numbers, $p_{i}=n_{i}/R_{i}~~(i=1,2)$,
and $\ln\delta=2\psi(1)-\psi(\lambda)-\psi(1-\lambda)$ for the digamma
function $\psi(x)$. Here, it is important to notice that the correlation
function depends on the parameters $k$ and $l$ only through the
normalization factor, $C_{-{\bm{p}}}^{k,l}$. Given this, we find
that the inner product $\big<\tilde{T}_{\bm{p}},\,\phi_{1}^{k,k}\big>$
depends on $k$ as follows, 
\begin{eqnarray}
 \big<\tilde{T}_{\bm p},\,\phi_1^{k,k}\big>=
  C_{-{\bm p}}^{k,\,k}\times \cdots
=e^{-i\frac{2\pi R_2\,k}{N}p^2}\times \cdots.
\end{eqnarray}
Therefore, it turns out that the tachyon profile $t^{k}({\bm{x}})$
of $\Psi_{0}^{k}$ satisfies the quasi-periodic relation: 
\begin{eqnarray}
  t^k\left(x^1,\,x^2+\frac{2\pi R_2}{N}\right)=
   t^{k-1}\left(x^1,\,x^2\right).
\label{eq:tk2}
\end{eqnarray}
Moreover, since the factor $C_{-{\bm{p}}}^{k,k}$ includes
$\delta_{0,-n_{1}\,({\rm mod}N)}$ we find
\begin{eqnarray}
   t^k\left(x^1+\frac{2\pi R_1}{N},\,x^2\right)=
   t^{k}\left(x^1,\,x^2\right).
\label{eq:tk1}
\end{eqnarray}

Similarly to the tachyon case, other profiles can be calculated also
by using a 3-point functions of the BCC operators. The matter vertex
operators are of the form $\partial^{n}X\cdots e^{ipX}$. Since the
operators $U,V$ are not included in the derivatives of $X^{1},X^{2}$\cite{Ishibashi:2016xak},
the 3-point function depends on the parameters $k$ and $l$ only
through $C_{-{\bm{p}}}^{k,l}$. Consequently, we find that all profiles
of the solution satisfy the quasi-periodic relations similar to (\ref{eq:tk2})
and (\ref{eq:tk1}), namely, the space representation of the solution
satisfies 
\begin{eqnarray}
   \Psi_0^k\left(x^1+\frac{2\pi R_1}{N},\,x^2\right)&=&
   \Psi_0^{k}\left(x^1,\,x^2\right),
\\
   \Psi_0^k\left(x^1,\,x^2+\frac{2\pi R_2}{N}\right)&=&
   \Psi_0^{k-1}\left(x^1,\,x^2\right).
\label{eq:x2trans}
\end{eqnarray}

Notice that the quasi-periodicity relations of these forms arise because
we have taken the BCC operators corresponding to the eigenstates of
$V=\exp(ix^{2}/R_{2})$ \cite{Ishibashi:2016xak}. If we take the
BCC operators corresponding to the eigenstates of $U=\exp(ix^{1}/R_{1})$,
we have a set of classical solutions which is periodic in the $x^1$
direction and quasi-periodic in the $x^2$ direction.
In addition, we should comment that, due to the
translational symmetry of the theory, we can generate other set of
solutions from $\Psi_0^k$ by arbitrary
displacement of the torus. In our case, we
simply just choose $\Psi_0^k$ as $|N|$ independent solutions in the sense
that they correspond to degenerate states in magnetic fields,
so-called Landau level.\footnote{
In a naive expectation,
the $|N|$ independent solutions $\Psi_0^k\ (k=1,\cdots,|N|)$ are physically
equivalent as a result of the translational symmetry and the relation
(\ref{eq:x2trans}).
However, it is difficult to connect the solutions to each other by a
gauge transformation, because we suffer from associativity 
anomalies\cite{HS} if the translation is represented as a gauge
transformation in open string field theory.
} 

\subsection{Tachyon profile}

Now let us calculate the tachyon profile. The inner product $\big<\tilde{T}_{\bm{p}},\,\phi_{1}^{k,l}\big>$
is rewritten in terms of correlation functions with the help of $KBc$
algebra: 
\begin{eqnarray}
 \big<\tilde{T}_{\bm p},\,\phi_1^{k,l}\big>
&=&
\left(\frac{2}{\pi}\right)^{\alpha'{\bm p}^2-1}\frac{1}{(2\pi)^2 R_1R_2}
\frac{1}{2}
\int_0^\infty \frac{dt_3}{\sqrt{\pi t_3}}
\int_0^\infty dt_2
\int_0^\infty \frac{dt_1}{\sqrt{\pi t_1}}e^{-t_1-t_2-t_3}
\times
\nn
&&\times\left\{\Big<e^{-i{\bm p}\cdot {\bm X}}
\left(L-\frac{1}{2}\right)
\partial \sigma^k(t_1+t_2) \bar{\sigma}^l(t_1)\Big>_L
\Big<c\partial c\left(L-\frac{1}{2}\right)
c(t_1+t_2) \Big>_L
\right.
\nn
&&-\left.\Big<e^{-i{\bm p}\cdot {\bm X}}
\left(L-\frac{1}{2}\right)
\sigma^k(t_1+t_2) \partial \bar{\sigma}^l(t_1)\Big>_L
\Big<c\partial c\left(L-\frac{1}{2}\right)
c(t_1) \Big>_L
\right\},
\label{profile_tacphi1}
\end{eqnarray}
where  $L=t_{1}+t_{2}+t_{3}+1$ and $\left<\cdots\right>_{L}$
denotes a correlation function on the infinite cylinder of circumference
$L$. The correlation functions which appear in (\ref{profile_tacphi1})
can be obtained by conformally transforming  (\ref{eq:3pttac}) and
the ghost correlation functions which are defined on the complex plane. We
eventually get the tachyon profile for the $|N|$-th solution
$\Psi_{0}^{|N|}$ as 
\begin{eqnarray}
 t({\bm x})&=& \left(1-\frac{1}{|\cos\pi\lambda|}\right)t_0
\nn
&&+\frac{2}{|\cos\pi\lambda|}
\Bigg[\sum_{m=1}^\infty
G_t\Big(\Big(\frac{Nm}{R_1}\Big)^2\,\Big)\cos\frac{Nmx^1}{R_1}
+\sum_{n=1}^\infty
G_t\Big(\Big(\frac{n}{R_2}\Big)^2\,\Big)\cos\frac{nx^2}{R_2}
\nn
&&
+2\sum_{n,m=1}^\infty G_t\Big(\Big(\frac{Nm}{R_1}\Big)^2
+\Big(\frac{n}{R_2}\Big)^2\Big)\,\Big)(-1)^{mn}
\cos\frac{Nmx^1}{R_1}\cos\frac{nx^2}{R_2}\,\Bigg],
\label{eq:tprofile_series}
\end{eqnarray}
where $t_0=\big<\tilde{T}_{\bm p},\,\Psi_{\rm tv}\big>$. The function
$G_t(u)$ is defined by
\begin{eqnarray}
 G_t(u)&=&-\frac{\alpha'u}{4}g(\alpha'u-1)\exp\left(-\frac{\alpha'u}{2}
(2\psi(1)-\psi(\lambda)-\psi(1-\lambda)-2\log 4)\right),
\label{eq:G_t(u)}
\\
 g(z)&=&
 \int_0^\infty dt \int_0^\infty ds \int_{s+t+1}^\infty dL
\frac{e^{1-L}}{\pi \sqrt{ts}}\left(
\frac{\sin\frac{\pi(s+t+1)}{L}}{2L\sin \frac{\pi(s+1/2)}{L}
\sin\frac{\pi(t+1/2)}{L}}\right)^z.
\label{eq:gz}
\end{eqnarray}
The tachyon profile for the $k$-th solution can be derived by using
(\ref{eq:tk2}).

\begin{figure}
\centerline{\includegraphics[width=7cm]{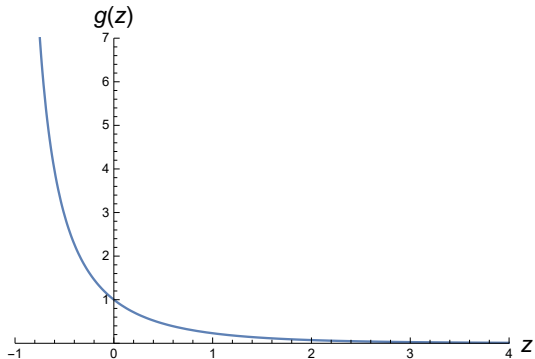}}
\caption{The plots of $g(z)$.}
\label{fig:1}
\end{figure}

Now, we carry out numerical evaluation of the tachyon profile. In
order to do so, we need to evaluate the triple integration on the
right hand side of (\ref{eq:gz}). However, this expression is inappropriate
for numerical integration, because the rate of convergence is very
slow due to the infinite integration region and it gives an inaccurate
value even by use of Monte Carlo method. An expression of $g\left(z\right)$
which is better suited than (\ref{eq:gz}) for numerical integration
can be obtained by a change of variables: 
\begin{eqnarray}
 g(z)&=&
\frac{2}{\pi}\int_0^1 dv
\int_0^{1-v}dw \int_0^{\frac{\pi}{2}} d\theta
e^{1-\frac{1}{v}}v^{z-3}\times
\nn
&&
\times
\left(\frac{\sin\pi w}{
2\sin\pi\left(\frac{v}{2}+(1-w-v)\sin^2\theta\right)
\sin\pi\left(\frac{v}{2}+(1-w-v)\cos^2\theta\right)
}\right)^z.
\end{eqnarray}
The plot of $g(z)$ obtained by numerically evaluating it using this
expression is depicted in Fig.~\ref{fig:1}. We find that the function
rapidly decreases as we increase $z$. The discontinuity of the profile
depends on the behavior of $g\left(z\right)$ when $z$ is very large.

\begin{figure}[h]
\centerline{\includegraphics[width=7cm]{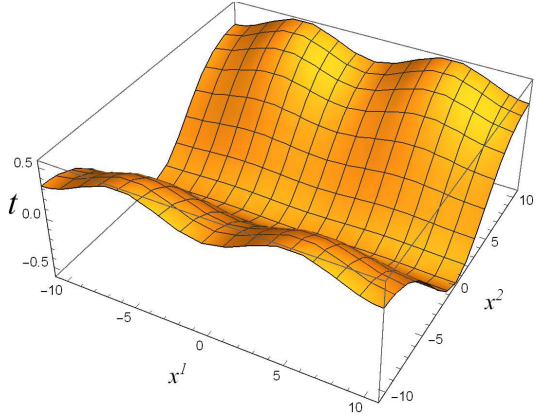}}
\caption{The numerical plots of the tachyon profile in the case of
 $N=2$ and $R_1=R_2=2\sqrt{3}$. Here, we take $\alpha'=1$.}
\label{fig:tprofile}
\end{figure}

The Fourier coefficients of (\ref{eq:tprofile_series}) are evaluated
by using the numerical results of $g(z)$ and the tachyon profile
can be obtained numerically by summing up the Fourier series. We find
that the Fourier coefficient approaches zero fast enough as $n,\,m\rightarrow\infty$
so that we can approximate the series by a finite summation over lower
modes. The plot of a result is depicted in
Fig.~\ref{fig:tprofile}.\footnote{The reason why the $x^1$ and $x^2$
dependence are different from each other is that
the solution is constructed by using the BCC operators corresponding to
the eigenstates of $V=\exp(ix^2/R_2)$.}

\subsection{Vector profile}

Next, we consider the vector profile of the magnetic solution. Since
the vector profile of $\Psi_{{\rm tv}}$ vanishes, the momentum space
vector profile of $\Psi_{0}^{k}$ is given by 
\begin{eqnarray}
  {A_{\bm p}}^\mu 
&=& \big<{\tilde{V}_{\bm p}}^\mu,\,\phi_1^{k,k}\big>-
\big<Q_{\rm B}{\tilde{V}_{\bm p}}^\mu,\,\phi_2^{k,k}\big>.
\label{profilevec}
\end{eqnarray}
As in the case of the tachyon profile, 
each inner product can be rewritten in terms of correlation functions on
infinite cylinder of circumference $L=t_1+t_2+t_3+1$:
\begin{eqnarray}
 \big<{\tilde{V}_{\bm p}}^\mu,\,\phi_1^{k,l}\big>
&=&
\left(\frac{2}{\pi}\right)^{\alpha'{\bm p}^2}\frac{1}{(2\pi)^2R_1R_2}i
\sqrt{\frac{2}{\alpha'}}\frac{1}{2}
\int_0^\infty \frac{dt_3}{\sqrt{\pi t_3}}
\int_0^\infty dt_2
\int_0^\infty \frac{dt_1}{\sqrt{\pi t_1}}e^{-t_1-t_2-t_3}
\times
\nn
&&\times\left\{\Big<\partial X^\mu e^{-i{\bm p}\cdot {\bm X}}
\left(L-\frac{1}{2}\right)
\partial \sigma^k(t_1+t_2) \bar{\sigma}^l(t_1)\Big>_L
\Big<c\partial c\left(L-\frac{1}{2}\right)
c(t_1+t_2) \Big>_L
\right.
\nn
&&-\left.\Big<\partial X^\mu e^{-i{\bm p}\cdot {\bm X}}
\left(L-\frac{1}{2}\right)
\sigma^k(t_1+t_2) \partial \bar{\sigma}^l(t_1)\Big>_L
\Big<c\partial c\left(L-\frac{1}{2}\right)
c(t_1) \Big>_L
\right\},
\label{profile_phi1}
\end{eqnarray}
\begin{eqnarray}
 \big<Q_{\rm B}{\tilde{V}_{\bm p}}^\mu,\,\phi_2^{k,l}\big>
&=&-\left(\frac{2}{\pi}\right)^{\alpha'{\bm p}^2}
\frac{1}{(2\pi)^2R_1R_2}i
\sqrt{\frac{\alpha'}{2}}ip^\mu \frac{1}{2}
\int_0^\infty \frac{dt_3}{\sqrt{\pi t_3}}
\int_0^\infty dt_2
\int_0^\infty \frac{dt_1}{\sqrt{\pi t_1}}e^{-t_1-t_2-t_3}
\times
\nn
&&\times\left\{\Big<e^{-i{\bm p}\cdot {\bm X}}
\left(L-\frac{1}{2}\right)
\partial \sigma^k(t_1+t_2) \bar{\sigma}^l(t_1)\Big>_L
\Big<c\partial c\partial^2 c\left(L-\frac{1}{2}\right)
c(t_1+t_2) B\Big>_L
\right.
\nn
&&+\left.\Big<e^{-i{\bm p}\cdot {\bm X}}
\left(L-\frac{1}{2}\right)
\sigma^k(t_1+t_2) \partial \bar{\sigma}^l(t_1)\Big>_L
\Big<c\partial c\partial^2 c\left(L-\frac{1}{2}\right)
Bc(t_1) \Big>_L
\right\},
\label{profile_phi2}
\end{eqnarray}
where we have used the expressions for the dual vector vertex operator,
(\ref{eq:Vec}) and (\ref{eq:QV}).

The computation of the correlation functions on the right hand side
of (\ref{profile_phi1})
is not so simple
because the vector vertex operator in (\ref{profile_phi1}) is not
a primary field. As derived in Appendix \ref{sec:conformaltrans},
under the conformal transformation $z=f(\xi)$, the vector vertex
transforms as follows; 
\begin{eqnarray}
\partial X^\mu e^{-i{\bm p}\cdot {\bm X}}(z)
=\left(\frac{d\xi}{dz}\right)^{\alpha'{\bm p}^2+1}
\left\{\partial X^\mu e^{-i{\bm p}\cdot {\bm X}}(\xi)
+\frac{i}{2}\alpha'p^\mu \frac{\frac{d^2\xi}{dz^2}}{
\left(\frac{d\xi}{dz}\right)^2}
e^{-i{\bm p}\cdot {\bm X}}(\xi)
\right\}.
\label{CTvec}
\end{eqnarray}
Thus, we need correlation functions for vector and tachyon vertex
operators to derive the vector profile. A detailed derivation of these
correlation functions is presented in Appendix~\ref{appendix:vecprofile}. Using
the results, it turns out that only the term (\ref{profile_phi1})
contributes and we find: 
\begin{eqnarray}
{A_{\bm p}}^\mu 
&=&
\frac{\lambda-\frac{1}{2}}{|\cos\pi\lambda|}\,\sqrt{2\alpha'}
\,i\epsilon^{\mu\nu} p_\nu\,C_{-{\bm p}}^{k,k}
\times
\nn
&&\times\,(-h)\,
\int_0^\infty \frac{dt_3}{\sqrt{\pi t_3}}\int_0^\infty dt_2
\int_0^\infty \frac{dt_1}{\sqrt{\pi t_1}}e^{-L+1}
\Bigg(\frac{2}{L}
\frac{\sin \theta_{t_2}}{\sin \theta_{t_1+\frac{1}{2}}
\sin \theta_{t_3+\frac{1}{2}}}\Bigg)^{h-1},
\label{eq:VecProfileMom}
\end{eqnarray}
where $h=\alpha'{\bm{p}}^{2}+1$ and $\theta_{s}$ is defined by $\theta_{s}=\pi s/L$. By the Fourier
transformation of (\ref{eq:VecProfileMom}), the position space representation
of the vector profile is obtained as follows: 
\begin{eqnarray}
 A_1({\bm x})&=&
\frac{\lambda-\frac{1}{2}}{|\cos\pi \lambda|}2\sqrt{2\alpha'}
\Bigg[
\sum_{n=1}^\infty \frac{n}{R_2}
G\Big(\Big(\frac{n}{R_2}\Big)^2\Big)
\sin\frac{nx^2}{R_2}
\nn
&&
+2\sum_{n,m=1}^\infty \frac{n}{R_2}
G\Big(\Big(\frac{Nm}{R_1}\Big)^2+
\Big(\frac{n}{R_2}\Big)^2\Big)(-1)^{mn}
\cos\frac{Nmx^1}{R_1}\sin\frac{nx^2}{R_2}
\Bigg],
\\
 A_2({\bm x})&=&
-\frac{\lambda-\frac{1}{2}}{|\cos\pi \lambda|}2\sqrt{2\alpha'}
\Bigg[
\sum_{n=1}^\infty \frac{Nm}{R_1}
G\Big(\Big(\frac{n}{R_2}\Big)^2\Big)
\sin\frac{Nmx^1}{R_1}
\nn
&&
+2\sum_{n,m=1}^\infty \frac{Nm}{R_1}G\Big(\Big(\frac{Nm}{R_1}\Big)^2+
\Big(\frac{n}{R_2}\Big)^2\Big)(-1)^{mn}
\sin\frac{Nmx^1}{R_1}\cos\frac{nx^2}{R_2}
\Bigg],
\end{eqnarray}
where the function $G(u)$ is defined in terms of $g\left(z\right)$
in (\ref{eq:gz}) as 
\begin{eqnarray}
 G(u)&=&(\alpha'u+1)\,g(\alpha'u)\exp\left(-\frac{\alpha'u}{2}
(2\psi(1)-\psi(\lambda)-\psi(1-\lambda)-2\log 4)\right).
\label{eq:G(u)}
\end{eqnarray}

As in the tachyon profile, we can numerically calculate the vector
profile by using the numerical results of $g(z)$ and summing up the
Fourier series. The difference of the two profiles is in the fact
that the Fourier coefficients of the vector profile include momentum
factors, $Nm/R_{1}$ or $n/R_{2}$. Hence the asymptotic behavior
of the profile may change in the ultraviolet region and discontinuities
could be found for the vector profile.

\begin{figure}
\centerline{\includegraphics[width=7cm]{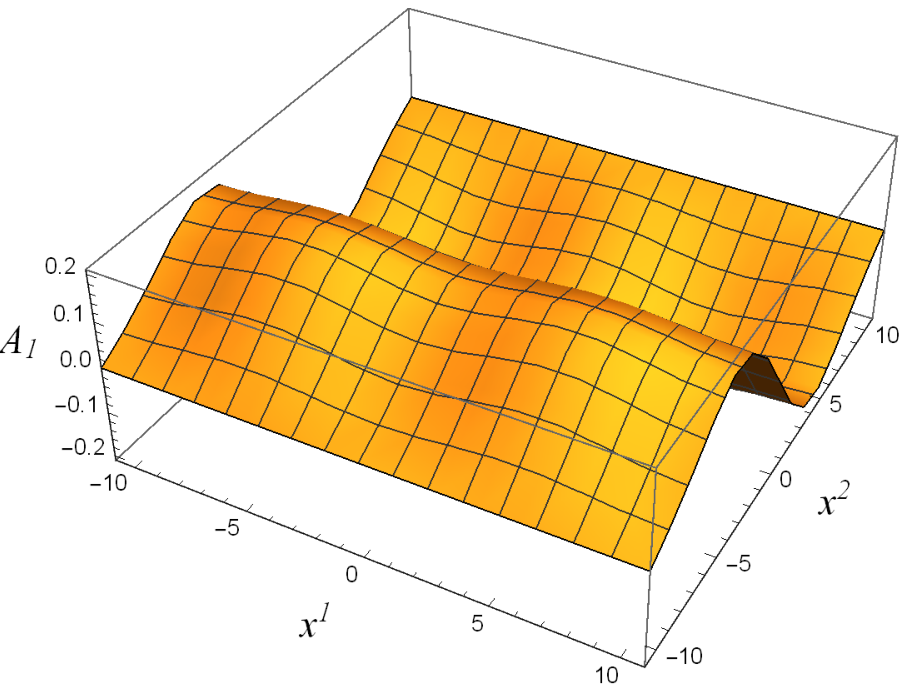}}
\caption{The numerical plots of the vector profile of $A_1({\bm x})$ in
the case of
$N=2$ and $R_1=R_2=2\sqrt{3}$. Here, we take $\alpha'=1$.}
\label{fig:3}
\end{figure}

\begin{figure}
\centerline{\includegraphics[width=7cm]{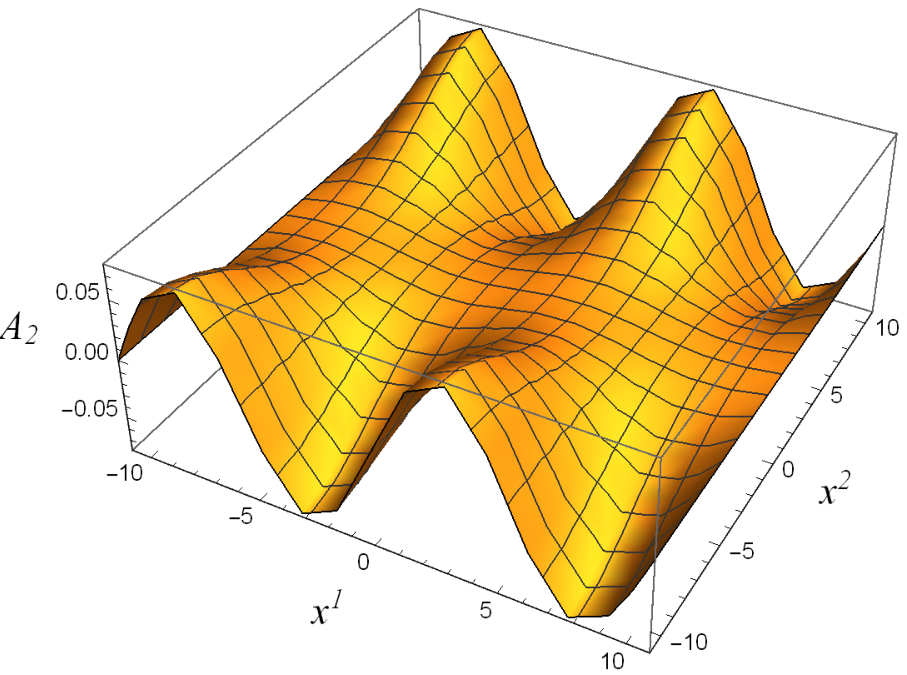}}
\caption{The numerical plots of the vector profile of
$A_2({\bm x})$ in the case of 
$N=2$ and $R_1=R_2=2\sqrt{3}$. Here, we take $\alpha'=1$.}
\label{fig:4}
\end{figure}

However, we observe that the Fourier coefficients for the vector profile
rapidly converges to zero as $n,m\to\infty$ and the position space
representation of the profile seems to be absolutely convergent. Consequently,
we get the plot shown in Figs.~\ref{fig:3} and \ref{fig:4}, and
we conclude that the vector profile has no
discontinuities.\footnote{From the vector profile, we can calculate the
quantity $\tilde{F}_{12}\equiv \partial_1 A_2-\partial_2 A_1$. The resulting
$\tilde{F}_{12}$ is given not as a constant but as a smooth function,
and the space average of $\tilde{F}_{12}$ over the torus becomes zero.
Since $\tilde{F}_{12}$ is not invariant under
gauge transformations in string field theory, it is no wonder that
$\tilde{F}_{12}$ does not correspond to the constant magnetic field.
}

Here, we would like to comment on profiles for other massive modes.
The reason why the tachyon and vector profiles are not discontinuous
is because the momentum space profile becomes zero rapidly for large
momenta. This behavior is due to the exponential factor $\delta^{-\alpha'{\bm{p}}^{2}/2}$
in the normalization factor of the 3 point function (\ref{eq:Cp}).\footnote{
It gives exponential factor  in (\ref{eq:G_t(u)}) and (\ref{eq:G(u)}), where
$2\psi(1)-\psi(\lambda)-\psi(1-\lambda)-2\log 4\ge 0$ for $0\le \lambda <1$.}
Since this exponential factor always appears in the expression of
other profiles, we expect that other profiles also do not have any
discontinuities.

\section{Gauge invariant observables for the classical solution}

\subsection{Calculation of gauge invariant observables}

Let us consider the gauge invariant observable (\ref{eq:ginv}) for
the the $k$-th solution (\ref{eq:kthsol}) with the following closed
string vertex operators at zero momentum: 
\begin{eqnarray}
 B(z,\bar{z})
  &\equiv&i(\partial X\bar{\partial}\tilde{X}
  -\partial \tilde{X}\bar{\partial}X)(z,\bar{z}),
\label{eq:B}
  \\
  G(z,\bar{z})
   &\equiv&(\partial X\bar{\partial}\tilde{X}
   +\partial \tilde{X}\bar{\partial}X)(z,\bar{z}),
\label{eq:H}
\end{eqnarray}
where $X=(X^1+iX^2)/\sqrt{2}$ and $\tilde{X}=(X^1-iX^2)/\sqrt{2}$.
These correspond to the antisymmetric tensor field $B_{12}$ and the
sum of graviton field $G_{11}+G_{22}$ in the spatial directions $X^{1}$
and $X^{2}$. Substituting the $k$-th solution (\ref{eq:kthsol})
into (\ref{eq:ginv}) and expressing it in terms of the correlation
functions on the infinite cylinder, the observables can be rewritten
as 
\begin{eqnarray}
 O_V(\Psi_0^k)&=&\left(1-\frac{1}{|\cos\pi\lambda|}
\right)O_V(\Psi_{\rm tv})
\nn
&&  -\int_0^\infty \frac{dt_1}{\sqrt{\pi t_1}}\int_0^\infty ds
  \int_0^\infty \frac{dt_2}{\sqrt{\pi t_2}} e^{-s-t_1-t_2}
  \nn
  &&\times \Big<V(s+t_1+i\infty,s+t_1-i\infty)\partial \sigma^k(s)
  \bar{\sigma}^k(0)\Big>_{C_L}
  \nn
  &&\times \Big<c(s+t_1+i\infty)c(s+t_1-i\infty)c(s)\Big>_{C_L},
\label{eq:OV}
\end{eqnarray}
where $L=s+t_{1}+t_{2}$ . For $B(z,\bar{z})$ and $G(z,\bar{z})$,
$O_{V}(\Psi_{{\rm tv}})$ has been calculated as\footnote{As in \cite{Kawano:2008ry}, if $V(z,\bar{z})$ is decomposed by the
matter primary field $V_{n}(z)$ as $V(z,\bar{z})=\sum_{m,n}\zeta_{mn}V_{m}(z)V_{n}(\bar{z})$,
and the OPE of the primary fields is $V_{m}(z)V_{n}(z')\sim v_{mn}/(z-z')^{2}$,
the observable for the tachyon vacuum solutions is given by 
\begin{eqnarray*}
 O_V(\Psi_{\rm tv})=\frac{1}{2\pi i}\sum_{m,n}\zeta_{mn}v_{mn}
\times {}_{\rm mat}\big<0\big|0\big>{}_{\rm mat}.
\end{eqnarray*}
}
\begin{eqnarray}
 &&
  O_B(\Psi_{\rm tv})=0,
  \\
 &&
  O_G(\Psi_{\rm tv})
  =-\frac{\alpha'}{2\pi i}\times (2\pi)^2R_1 R_2.
\end{eqnarray}
Here we normalize the observable by dividing it by the volume of the
directions other than $X^{1}$ and $X^{2}$.

For the vertex operators (\ref{eq:B}) and (\ref{eq:H}), the matter
correlation function in (\ref{eq:OV}) can be derived from the correlator
\begin{eqnarray}
 \big<\bar{\sigma}^k\big|
  \partial X(z)\bar{\partial}\tilde{X}(\bar{w})\big|\sigma^l\big>
  &=& -\frac{\alpha'}{2}
  \frac{\lambda \left(\frac{z}{\bar{w}}\right)^{1-\lambda}
  +(1-\lambda)\left(\frac{\bar{w}}{z}\right)^\lambda}{(z-\bar{w})^2}
  \times \big<\bar{\sigma}^k\big|\sigma^l\big>,
\label{eq:2ptfunc}
\end{eqnarray}
and its conjugate. Here $\big<\bar{\sigma}^{k}\big|\sigma^{l}\big>$
is given by
\begin{eqnarray}
 \big<\bar{\sigma}^k\big|\sigma^l\big>
=\frac{(2\pi)^2R_1R_2}{|\cos\pi\lambda|}\delta_{k,l}.
\end{eqnarray}
The matter 4-point function in (\ref{eq:OV}) on the infinite cylinder
can be obtained from (\ref{eq:2ptfunc}) by a conformal transformation.
To make our calculation well-defined, we regularize the correlators
by replacing $\pm i\infty$ by $\pm iM$ and taking the limit $M\to\infty$.
As a result, we find 
\begin{eqnarray}
 &&
  \big<
  \partial X(x+iM)\bar{\partial}\tilde{X}(x-iM)
  \sigma^k(x_1)\bar{\sigma}^k(x_2)\big>_{C_L}
  \nn
  &\rightarrow&
  -\frac{\alpha'}{2}\left(\frac{\pi}{L}\right)^2
  (-4)e^{-\frac{\pi}{L}4M}
  \left\{
   \lambda e^{i\frac{2\pi}{L}(1-\lambda)(x_1-x_2)}
   +(1-\lambda) e^{-i\frac{2\pi}{L}\lambda(x_1-x_2)}
	  \right\}
  \times\frac{(2\pi)^2R_1R_2}{|\cos\pi\lambda|}.
\end{eqnarray}
 Multiplying the ghost correlator, the 4-point function in (\ref{eq:OV})
is given as 
\begin{eqnarray}
 &&
  \big<
  \partial X(s+t_1+i\infty)\bar{\partial}\tilde{X}(s+t_1-i\infty)
  \partial \sigma^k(s)\bar{\sigma}^k(0)\big>_{C_L}
    \big<
  c(s+t_1+i\infty)c(s+t_1-i\infty)c(s)\big>_{C_L}
  \nn
  &=&
  \frac{\alpha'}{2}\lambda(1-\lambda)
  \left\{
   e^{-i\frac{2\pi}{L}\lambda s}
   -e^{i\frac{2\pi}{L}(1-\lambda)s}
	  \right\}
  \times \frac{(2\pi)^2R_1R_2}{|\cos\pi\lambda|},
\label{eq:correlfunc}
\end{eqnarray}
where $L=s+t_1+t_2$. Similarly its conjugate is given by
\begin{eqnarray}
 &&
  \big<
  \bar{\partial} X(s+t_1-i\infty)\partial\tilde{X}(s+t_1+i\infty)
  \partial \sigma^k(s)\bar{\sigma}^k(0)\big>_{C_L}
    \big<
  c(s+t_1+i\infty)c(s+t_1-i\infty)c(s)\big>_{C_L}
  \nn
  &=&
  \frac{\alpha'}{2}\lambda(1-\lambda)
  \left\{
   e^{-i\frac{2\pi}{L}(1-\lambda) s}
   -e^{i\frac{2\pi}{L}\lambda s}
	  \right\}
  \times
  \frac{(2\pi)^2R_1R_2}{|\cos\pi\lambda|},
\end{eqnarray}
which is also obtained by acting the $X^{2}$ parity transformation
$\lambda\leftrightarrow1-\lambda$ on (\ref{eq:correlfunc}) \cite{Ishibashi:2016xak}.

Having calculated the correlation functions, 
the observables can be obtained by integration of (\ref{eq:OV}):
\begin{eqnarray}
 O_B(\Psi_0)
    &=&
    \frac{\alpha'}{\pi i}\times (2\pi)^2 R_1 R_2 \times
    \frac{2\pi\alpha'F_{12}}{\sqrt{1+(2\pi\alpha' F_{12})^2}},
\label{eq:OB}
\\
 O_G(\Psi_0)
    &=&\frac{i \alpha'}{2\pi}(2\pi)^2R_1R_2
    \times
    \nn
    &&
    \times
    \left\{
					       1-\sqrt{1+(2\pi\alpha'F_{12})^2}
    +2\bigg(\sqrt{1+(2\pi\alpha'F_{12})^2}-
	 \frac{1}{\sqrt{1+(2\pi\alpha'F_{12})^2}}\bigg)\right\}.
\label{eq:OG}
\end{eqnarray}

\subsection{Comparison with Dirac-Born-Infeld action}

Let us check if the gauge invariant observables (\ref{eq:OB}) and
(\ref{eq:OG}) are consistent with what is known about the magnetic
background. Here we consider a D2-brane extended in the $X^{1}$ and
$X^{2}$ directions for simplicity. The coupling of the D2-brane to
massless modes of closed strings is described by the Dirac-Born-Infeld
(DBI) action:\footnote{
Hereafter, we adopt the static gauge and normalize the action by dividing it by the volume of the time direction.
}
\begin{eqnarray}
 S&=&-T_2  e^{-\Phi} (2\pi)^2 R_1R_2
\sqrt{\det(G_{ab}+B_{ab}+2\pi\alpha'F_{ab})}, 
\end{eqnarray}
where $T_{2}$ is the D2-brane tension, and $\Phi$, $G_{ab}$ and
$B_{ab}$ denote the induced fields on the brane. 

Fixing to the static gauge, we find the variation of the DBI action
under an infinitesimal variation of $B_{12}$ around the flat background,
namely $\left<G_{ab}\right>=\delta_{ab}$ and $\left<B_{ab}\right>=0$:
\begin{eqnarray}
 \delta S&=&
-T_2   e^{-\Phi}
(2\pi)^2 R_1R_2
  \frac{2\pi\alpha'F_{12}}{\sqrt{1+(2\pi\alpha'F_{12})^2}}\delta
  B_{12}.
\end{eqnarray}
The observable (\ref{eq:OB}) can be expressed by using the DBI action
as 
\begin{eqnarray}
O_B(\Psi_0)=\frac{i \alpha'}{\pi T_2e^{-\Phi}}
\Bigg\{
\frac{\delta S}{\delta B_{12}}\Big|_{F_{12}\neq 0}
-\frac{\delta S}{\delta B_{12}}\Big|_{F_{12}= 0}
\Bigg\}.
\label{eq:OBdS}
\end{eqnarray}

Similarly, we find the variation of the DBI action under an infinitesimal
variation of $G_{ab}$: 
\begin{eqnarray}
 \delta S
&=&
-T_2   e^{-\Phi}
(2\pi)^2 R_1R_2 \times
\nn
&&
\hspace{-.5cm}\times\frac{1}{2}\left\{
\frac{1}{2}\sqrt{1+(2\pi\alpha' F_{12})^2}
-\sqrt{1+(2\pi\alpha' F_{12})^2}
+\frac{1}{\sqrt{1+(2\pi\alpha' F_{12})^2}}
\right\}(\delta G_{11}+\delta G_{22}).
\label{eq:dSG}
\end{eqnarray}
Notice that the variation $\delta G_{12}$ does not appear in this
result and the graviton field is included implicitly in the dilaton
field as $\Phi=\Phi_{24}+(1/4)\ln\det G_{ab}$ when the space-time
is of the form $M^{24}\times T^{2}$ \cite{PolchinskiI}. From (\ref{eq:dSG})
we get the relation 
\begin{eqnarray}
 O_G(\Psi_0)
=\frac{i \alpha'}{\pi T_2e^{-\Phi}}
\Bigg\{
\left(
\frac{\delta S}{\delta G_{11}}
+\frac{\delta S}{\delta G_{22}}
\right)\Big|_{F_{12}\neq 0}
-
\left(
\frac{\delta S}{\delta G_{11}}
+\frac{\delta S}{\delta G_{22}}
\right)\Big|_{F_{12}= 0}
\Bigg\}.
\label{eq:OGdS}
\end{eqnarray}

These results (\ref{eq:OBdS}) and (\ref{eq:OGdS}) show that the
observables correctly reflect the coupling of the D2-brane with the
constant magnetic field to the closed string modes. Consequently,
it is explicitly found that (\ref{eq:kthsol}) can be regarded as
the classical solution corresponding to magnetic field condensation.

\subsection{T-dual of a D2-brane with $F_{12}\neq 0$.}

We would like to examine if the couplings (\ref{eq:OB}) and (\ref{eq:OG})
are consistent with T-duality. It is well known that the D2-brane
is transformed into a D-string tilted in the dual torus, which is
extended along the line\cite{PolchinskiI} 
\begin{eqnarray}
 {X'}^1=2\pi \alpha' F_{12} X^2+{\rm const.},
\end{eqnarray}
where ${X'}^{1}$ denotes the coordinate dual to $X^{1}$. From the
Dirac quantization condition (\ref{Dirac}), it follows that the D-string
winds $N$ times around the $X^{2}$ direction.

The DBI action for the D-string is given by
\begin{eqnarray}
 S&=&-T_1 \int d\xi\,
e^{-\Phi'}\sqrt{
\det{\left[
G'_{ab}+B'_{ab}+2\pi\alpha'F_{ab}
\right]}
},
\label{eq:Dstring}
\end{eqnarray}
where $T_{1}$ is the D-string tension and $G'_{ab}$, $B'_{ab}$
and $\Phi'$ are the induced fields in the dual space-time.
Here, we take the coordinate $\xi$ on the D-string to coincide with
$X^{2}$. Then, the embedding function $\phi(\xi)$ is given by 
\begin{eqnarray}
 \phi^1(\xi)=2\pi\alpha' F_{12} X^2 +{\rm const.},
~~~~\phi^2(\xi)=X^2.
\end{eqnarray}
In this gauge, the induced metric is calculated as
\begin{eqnarray}
 G'_{11}(\xi)=G'_{MN}\frac{\partial \phi^M}{\partial \xi}
\frac{\partial \phi^N}{\partial \xi}
=(2\pi\alpha'F_{12})^2G'_{11}+4\pi\alpha'F_{12}G'_{12}+G'_{22}.
\end{eqnarray}
Substituting this into the DBI action (\ref{eq:Dstring}), we can
calculate the variation of $S$ under an infinitesimal variation of
$G'_{ab}$: 
\begin{eqnarray}
 \delta S&=&-T_1 e^{-\Phi'}2\pi R_2\times
\nn
&&\times\bigg\{
-\frac{1}{4}\sqrt{1+(2\pi \alpha'F_{12})^2}
(\delta G'_{11}+\delta G'_{22})
\nn
&&
+\frac{1}{2}
\frac{1}{\sqrt{1+(2\pi\alpha'F_{12})^2}}
\left\{
(2\pi\alpha'F_{12})^2\delta G'_{11}+4\pi\alpha'F_{12}
\delta G'_{12}+\delta G'_{22}\right\}
\bigg\},
\end{eqnarray}
where we have used the relation $\Phi'=\Phi'_{24}-(1/4)\ln\det G'_{ab}$
as in the D2-brane case.

By using the relations between the parameters\cite{PolchinskiI} 
\begin{eqnarray}
 e^{-\Phi'}=\frac{R_1}{\sqrt{\alpha'}}e^{-\Phi},
~~~T_1=2\pi\sqrt{\alpha'}T_2, 
\end{eqnarray}
$\delta S$ can be
rewritten as
\begin{eqnarray}
  \delta S
&=&-T_2 e^{-\Phi}(2\pi)^2 R_1 R_2\times
\nn
&&\times\bigg\{
\frac{2\pi\alpha'F_{12}}{\sqrt{1+(2\pi\alpha'F_{12})^2}}\delta G'_{12}
\nn
&&
+\left(
-\frac{1}{4}\sqrt{1+(2\pi \alpha'F_{12})^2}
+\frac{1}{2}
\frac{1}{\sqrt{1+(2\pi\alpha'F_{12})^2}}
\right)
(-\delta G'_{11})
\nn
&&
+\left(
-\frac{1}{4}\sqrt{1+(2\pi \alpha'F_{12})^2}
+\frac{1}{2}
\frac{1}{\sqrt{1+(2\pi\alpha'F_{12})^2}}
\right)
\delta G'_{22}
\bigg\}.
\label{eq:dSD1}
\end{eqnarray}
Combining with the results (\ref{eq:OB}) and (\ref{eq:OG}), we find that
\begin{eqnarray}
O_B(\Psi_0)&=&
 \frac{i \alpha'}{\pi T_2e^{-\Phi}}
\Bigg\{
\frac{\delta S}{\delta G'_{12}}\Big|_{F_{12}\neq 0}
-\frac{\delta S}{\delta G'_{12}}\Big|_{F_{12}= 0}
\Bigg\},
\\
 O_G(\Psi_0)
&=&\frac{i \alpha'}{\pi T_2e^{-\Phi}}
\Bigg\{
\left(
\frac{\delta S}{\delta (-G'_{11})}
+\frac{\delta S}{\delta G'_{22}}
\right)\Big|_{F_{12}\neq 0}
-
\left(
\frac{\delta S}{\delta (-G'_{11})}
+\frac{\delta S}{\delta G'_{22}}
\right)\Big|_{F_{12}= 0}
\Bigg\}.
\end{eqnarray}
These relations implies that the gauge invariant observables reproduce
correct couplings of the D-sting to the closed string modes in the
dual space, because, in the dual space, the vertex operators ($\ref{eq:B}$)
and (\ref{eq:H}) correspond to $G'_{12}$ and a $-G'_{11}+G'_{22}$
respectively.

\section{Concluding remarks}

In this paper, we have studied and explicitly calculated the tachyon
and vector profiles for the constant magnetic field solution on a
torus constructed by following Erler-Maccaferri's method. In addition,
we have calculated gauge invariant observables for the solution and
found that the solution reproduces correct couplings of the D-brane
with a constant magnetic field to the closed string modes.

A remarkable feature of the resulting profiles is that they have no
discontinuity on the torus. This result does not seem to be consistent
with the fact that the solution corresponds to a topologically
nontrivial configuration. In low energy field theory, the gauge field
will have discontinuities if one tries to describe the configuration
without dividing the torus into patches. We expect that the same thing
happens in string field theory, but we have found that profiles for any
states do not have discontinuities. One possibility is that profiles for
normalized states are not the right quantities to be chosen in observing
such phenomena. Instead, we might have to consider a sum of infinitely
many profiles or even a nonlinear functional of the string
field. Another possibility is that our results could be an indication of
nonlocality of string field theory.
Since the star product is a nonlocal operation from
the target space viewpoint, even if a string field is defined in a
coordinate patch, it can spread beyond the boundary after a gauge
transformation.
Therefore, the non-existence of discontinuity itself may be a natural
consequence of the nonlocality of string field theory.

Another important question about the solution is how we can define
the topological invariant characterizing the constant magnetic field
solution in the framework of  string field theory. The magnetic
field is proportional to an integer due to the Dirac quantization
condition, which is derived based on low energy theory. This quantization
condition should be derived from the string field theory itself. Unfortunately,
the gauge invariant observables calculated in this paper do not provide
any clue about such a quantization condition, although it is interesting
to notice that the observable with the antisymmetric tensor mode becomes
non-zero as a result of the magnetic field background. In order to
capture the ``topological'' nature, we may need new insights from
noncommutative geometry, matrix theory and so on\cite{Grosse:1995jt,Connes:1997cr,Witten:2000cn,CarowWatamura:2004ct}.

There are many interesting possibilities for future exploration related
to the constant magnetic field solution. Suppose that we consider
the configuration which is T-dual to the magnetic field solution along
both $X^{1}$ and $X^{2}$ directions. The resulting configuration
may correspond to a configuration of multiple D-particles on the torus.
Such a system was studied in the case of a non-compact space \cite{Ishibashi:1998ni}.
In this case, the coordinates of the D-particles become noncommutative
and the observable (\ref{eq:OB}) represents the coupling of the D-particle
to a symplectic form characterizing the noncommutativity. Therefore,
D-particles with noncommutative coordinates may be described in terms
of the string field theory in this dual background. Such a string
field theory may be considered as another version of Matrix theories
\cite{Banks:1996vh,Ishibashi:1996xs} although it has no supersymmetry.
It is also possible to make the relation between noncommutative geometry
and constant magnetic field background manifest by a similarity transformation
for string fields\cite{Sugino:1999qd,Kawano:1999fw}. It will be interesting
to find such a similarity transformation in the background of the
magnetic field solution.

\section*{Acknowledgments}
The authors would like to thank T.~Erler, M.~Kudrna, C.~Maccaferri,
M.~Schnabl, Y.~Kaneko and S.~Watamura for helpful comments.
We also thank the organizers of ``SFT@HIT'' at Holon, especially
M.~Kroyter, for hospitality.
T.~M. also would like to thank the organizers of  
``String Field Theory and String Phenomenology'' 
at Harish-Chandra Research Institute, Allahabad, 
for their hospitality and for providing him a stimulating environment.
This work was supported in part by JSPS Grant-in-Aid
for Scientific Research (C) (JP25400242), JSPS Grant-in-Aid for Young
Scientists (B) (JP25800134), and
JSPS Grant-in-Aid for Scientific Research (C) (JP15K05056), and Nara
Women's University Intramural Grant for Project Research.
The research of T.M. was supported by the Grant Agency of the Czech
Republic under the grant 17-22899S.
The numerical computation in this work was partly carried out 
at the Yukawa Institute Computer Facility.
\appendix
\section{Conformal transformation of the vector vertex operator
\label{sec:conformaltrans}}

In this appendix, we would like to derive the conformal transformation
(\ref{CTvec}) of the vector vertex operator. The OPE of the vertex
operator with the energy momentum tensor is given by 
\begin{eqnarray}
 T(\zeta)\,\partial X^\mu e^{i{\bm p}\cdot {\bm X}}(\xi)
&\sim& \frac{-i\alpha' p^\mu}{(\zeta-\xi)^3}e^{i{\bm p}\cdot {\bm X}}(\xi)
+\frac{\alpha'{\bm p}^2+1}{(\zeta-\xi)^2}
\partial X^\mu e^{i{\bm p}\cdot {\bm X}}(\xi)
+
\frac{1}{\zeta-\xi}
\partial \left(
\partial X^\mu e^{i{\bm p}\cdot {\bm X}}\right)
(\xi).
\label{OPEvec}
\end{eqnarray}
The right hand side includes the tachyon vertex operator which is
a primary field of weight $\alpha'{\bm{p}}^{2}$: 
\begin{eqnarray}
 T(\zeta)\,e^{i{\bm p}\cdot {\bm X}}(\xi)
&\sim& 
\frac{\alpha'{\bm p}^2}{(\zeta-\xi)^2}
e^{i{\bm p}\cdot {\bm X}}(\xi)
+
\frac{1}{\zeta-\xi}
\partial \left(
e^{i{\bm p}\cdot {\bm X}}\right)
(\xi).
\label{OPEtac}
\end{eqnarray}

Let us consider a conformal transformation $\xi\to z=f(\xi)$, under
which a field $\phi(\xi)$ is transformed as 
\begin{eqnarray}
\phi(\xi)&\rightarrow& f\circ \phi(\xi) =U_f \phi(\xi) U_f^{-1}.
\end{eqnarray}
Here, the operator $U_{f}$ is given in terms of the energy-momentum
tensor as follows: 
\begin{eqnarray}
 U_f=e^{{\cal T}(v)},~~~~{\cal T}(v)=\oint \frac{d\xi}{2\pi
  i}v(\xi)T(\xi),~~~~
 f(\xi)=e^{v(\xi)\partial }\xi.
\end{eqnarray}
Since the tachyon vertex is primary, it is transformed as
\begin{eqnarray}
 U_f\,e^{i{\bm p}\cdot {\bm X}}(\xi)\,U_f^{-1}=
\left(\frac{dz}{d\xi}\right)^{\alpha'{\bm p}^2}
e^{i{\bm p}\cdot {\bm X}}(z).
\end{eqnarray}

From the OPEs (\ref{OPEvec}) and (\ref{OPEtac}), we can derive the
commutation relations of the vertex operators with ${\cal T}(v)$:
\begin{eqnarray}
 \big[{\cal T}(v),\,\partial X^\mu e^{i{\bm p}\cdot {\bm X}}(\xi)\big]
&=& -\frac{i}{2}p^\mu
(\partial^2 v)\,e^{i{\bm p}\cdot {\bm X}}(\xi)
\nn
&&~~~~~~~~~+
(\alpha'{\bm p}^2+1) (\partial v)\,\partial X^\mu
e^{i{\bm p}\cdot {\bm X}}(\xi)
+v\,\partial \big(\partial X^\mu 
e^{i{\bm p}\cdot {\bm X}}\big)(\xi),
\label{CRvec}
\\
 \big[{\cal T}(v),\,e^{i{\bm p}\cdot {\bm X}}(\xi)\big]
&=& \alpha'{\bm p}^2 (\partial v)\,e^{i{\bm p}\cdot {\bm X}}(\xi)
+v\,\partial e^{i{\bm p}\cdot {\bm X}}(\xi).
\label{CRtac}
\end{eqnarray}
Let us consider the transformation of the vector vertex operator under
the following one parameter family of conformal transformations: 
\begin{eqnarray}
 e^{t{\cal T}(v)}\,\partial X^\mu e^{i{\bm p}\cdot {\bm X}}(\xi)
\, e^{-t{\cal T}(v)}.
\end{eqnarray}
By using the commutation relations (\ref{CRvec}) and (\ref{CRtac}),
we can see that the result should be expressed as 
\begin{eqnarray}
 Y_t(\xi)\,\partial X^\mu e^{i{\bm p}\cdot {\bm X}}(\xi)
+Z_t^\mu(\xi)\,e^{i{\bm p}\cdot {\bm X}}(\xi),
\end{eqnarray}
where $Y_{t}(\xi)$ and $Z_{t}(\xi)$ are some functions of $t$ and
$\xi$. Differentiating these with respect to $t$ and using (\ref{CRvec})
and (\ref{CRtac}), we obtain the equations 
\begin{eqnarray}
 \frac{dY_t(\xi)}{dt}&=&(\alpha'{\bm p}^2+1)v'\big(f_t(\xi)\big)\,Y_t(\xi),
\nn
 \frac{dZ_t^\mu(\xi)}{dt}&=&(\alpha'{\bm p}^2)
 v'\big(f_t(\xi)\big)\,Z_t^\mu(\xi)
-\frac{i}{2}\alpha' p^\mu v''\big(f_t(\xi)\big)Y_t(\xi),
\end{eqnarray}
where $f_{t}(\xi)$ is defined by $f_{t}(\xi)=\exp(t\,v(\xi)\partial)\,\xi$.
Integrating these with the initial conditions $Y_{t=0}=1$ and $Z_{t=0}=0$,
we find 
\begin{eqnarray}
 Y_t(\xi)=\left(f_t'(\xi)\right)^{\alpha'{\bm p}^2+1},
~~~~Z_t^\mu(\xi)=-\frac{i}{2}p^\mu \frac{f''_t(\xi)}{f'_t(\xi)}
\left(f_t'(\xi)\right)^{\alpha'{\bm p}^2}.
\end{eqnarray}
By setting $t=1$ and $p^{\mu}\rightarrow-p^{\mu}$, we obtain the
formula (\ref{CTvec}).

\section{Calculation of correlation functions in the vector profile
\label{appendix:vecprofile}}

In this appendix, we will show how to calculate the correlation functions
which appear in (\ref{profile_phi1}) and (\ref{profile_phi2}). 

Here we begin with the calculation of the following 4-point functions
on the complex $\zeta$ plane: 
\begin{eqnarray}
 \Big<\partial X(\zeta)e^{-i{\bm p}\cdot {\bm X}}(\xi)
\sigma_*^k(\xi_1)\bar{\sigma}_*^l(\xi_2)\Big>.
\end{eqnarray}
Here $X=(X^{1}+iX^{2})/\sqrt{2}$, and $\xi$, $\xi_{1}$ and $\xi_{2}$
are taken to be real and satisfy $\xi_{2}<\xi_{1}<\xi$, so that the
operator $e^{-i{\bm{p}}\cdot{\bm{X}}}$ is on the boundary with the
Neumann boundary condition. The correlation function is a $1$-form
on the complex plane with respect to the variable $\zeta$ and it
behaves as
\renewcommand{\arraystretch}{1.3}
\begin{eqnarray}
 &\sim&
\left\{
\begin{array}{ll}
 (\zeta-\xi_1)^{-\lambda}&  (\zeta\sim \xi_1)\\
 (\zeta-\xi_2)^{-1+\lambda}& (\zeta\sim \xi_2)\\
 \zeta^{-2}& (\zeta\sim \infty).
\end{array}
\right.
\end{eqnarray}
\renewcommand{\arraystretch}{1}
The first two are derived directly from boundary conditions of $X$.
Moreover, in the limit $\zeta\sim\xi$, we find that the 4-point function
behaves as 
\begin{eqnarray}
=  \frac{i\alpha'p}{\zeta-\xi}\,
 \Big<e^{-i{\bm p}\cdot {\bm X}}(\xi)
\sigma_*^k(\xi_1)\bar{\sigma}_*^l(\xi_2)\Big>+{\rm regular\ terms},
\end{eqnarray} 
which fixes the normalization of the four point function.
These conditions determine the 4-point function as
\begin{eqnarray}
  \Big<\partial X(\zeta)e^{-i{\bm p}\cdot {\bm X}}(\xi)
\sigma_*^k(\xi_1)\bar{\sigma}_*^l(\xi_2)\Big>
&=&i\alpha'p\frac{(\zeta-\xi_1)^{-\lambda}(\zeta-\xi_2)^{-1+\lambda}
(\xi-\xi_1)^\lambda(\xi-\xi_2)^{1-\lambda}}{\zeta-\xi}
\nn
&&
 \times \Big<e^{-i{\bm p}\cdot {\bm X}}(\xi)
 \sigma_*^k(\xi_1)\bar{\sigma}_*^l(\xi_2)\Big>,
 \label{eq:4ptvec}
\end{eqnarray} 
where the 3-point function on the right hand side is given by (\ref{eq:3pttac})
\cite{Ishibashi:2016xak}. By taking the limit $\zeta\rightarrow\xi$ in
(\ref{eq:4ptvec}), we obtain a correlation function involving the
vector vertex operator on the complex plane: 
\begin{eqnarray}
   \Big<\partial X e^{-i{\bm p}\cdot {\bm X}}(\xi)
\sigma_*^k(\xi_1)\bar{\sigma}_*^l(\xi_2)\Big>
=i\alpha'p\frac{-\xi+(1-\lambda)\xi_1+\lambda \xi_2}{(\xi-\xi_1)(\xi-\xi_2)}
\frac{(\xi_1-\xi_2)^{\alpha'{\bm p}^2-\lambda(1-\lambda)}}{
(\xi-\xi_1)^{\alpha'{\bm p}^2}(\xi-\xi_2)^{\alpha'{\bm p}^2}}\,C_{-\bm
p}^{l,k}.
\label{3ptstar}
\end{eqnarray}
Taking the normalization of the modified BCC operators \cite{Ishibashi:2016xak}
into account, we obtain 
\begin{eqnarray}
&&
   \Big<\partial X e^{-i{\bm p}\cdot {\bm X}}(\xi)
\sigma^k(\xi_1)\bar{\sigma}^l(\xi_2)\Big>
\nn
&=&i\alpha'p\,C_{-\bm p}^{l,k}\,\frac{(2\pi)^2R_1R_2}{|\cos\pi\lambda|}\,
\{-\xi+(1-\lambda)\xi_1+\lambda \xi_2\}
\frac{(\xi_1-\xi_2)^{\alpha'{\bm p}^2-1}}{
(\xi-\xi_1)^{\alpha'{\bm p}^2}(\xi-\xi_2)^{\alpha' {\bm p}^2}}.
\label{3ptcomplex}
\end{eqnarray}
The correlation function for the conjugate vertex operator $\partial\tilde{X}e^{-i{\bm{p}}\cdot{\bm{X}}}\ (\tilde{X}=(X^{1}-iX^{2})/\sqrt{2})$ can
be calculated in a similar way and we find 
\begin{eqnarray}
&&
   \Big<\partial \tilde{X} e^{-i{\bm p}\cdot {\bm X}}(\xi)
\sigma^k(\xi_1)\bar{\sigma}^l(\xi_2)\Big>
\nn
&=&i\alpha'\tilde{p}\,
C_{-\bm p}^{l,k}\,\frac{(2\pi)^2R_1R_2}{|\cos\pi\lambda|}\,
\{-\xi+\lambda\xi_1+(1-\lambda) \xi_2\}
\frac{(\xi_1-\xi_2)^{{\alpha'{\bm p}^2}-1}}{
(\xi-\xi_1)^{\alpha'{\bm p}^2}(\xi-\xi_2)^{\alpha'{\bm p}^2}}.
\label{3ptcomplexdual}
\end{eqnarray} 
(\ref{3ptcomplexdual}) can also be obtained by using the fact that
the $X^{2}$ parity transformation $X^{2}\to-X^{2}$ corresponds to
the following transformations of the parameters:\cite{Ishibashi:2016xak}
\[
p\to\tilde{p},\qquad\lambda\to1-\lambda\,.
\]

Having found the 3-point functions on the complex plane, we can find
the one on the infinite cylinder with circumference $L$ by the conformal
transformation: 
\begin{eqnarray}
 z=\frac{L}{\pi}\arctan \xi.
\end{eqnarray}
By using the transformation law (\ref{CTvec}), we obtain the following
3-point functions on the cylinder from (\ref{3ptcomplex}) and (\ref{3ptcomplexdual}):
\begin{eqnarray}
&&
   \Big<\partial X^\mu e^{-i{\bm p}\cdot {\bm X}}(z_1)
\sigma^k(z_2)\bar{\sigma}^l(z_3)\Big>_L 
\nn
&=&
\left(\frac{\pi}{L}\right)^h C_{-{\bm p}}^{l,k}
\frac{(2\pi)^2R_1R_2}{|\cos\pi\lambda|}\,
\frac{\big\{ \sin \frac{\pi(z_2-z_3)}{L} \big\}^{h-1}}{\big\{
\sin \frac{\pi(z_1-z_2)}{L} \sin\frac{\pi(z_1-z_3)}{L}
\big\}^h }\times
\nn
&&\times\Bigg[\Big(\lambda-\frac{1}{2}\Big)
\alpha'\epsilon^{\mu\nu}
p_\nu \sin\frac{\pi(z_2-z_3)}{L}
\nn
&&~~~~~~-\frac{i}{2}\alpha'p^\mu\bigg\{
\sin\frac{\pi(z_1-z_2)}{L}\cos\frac{\pi(z_1-z_3)}{L}
+\sin\frac{\pi(z_1-z_3)}{L}\cos\frac{\pi(z_1-z_2)}{L}
\bigg\}\Bigg],
\label{3ptfunc_vec}
\end{eqnarray}
where $h=\alpha'{\bm{p}}^{2}+1$ and $\epsilon^{12}=-\epsilon^{21}=1$.

The matter correlation functions that appear in (\ref{profile_phi1})
are calculated by using (\ref{3ptfunc_vec}) as
\begin{eqnarray}
&&
\Big<\partial X^\mu e^{-i{\bm p}\cdot {\bm X}}
\left(L-\frac{1}{2}\right)
\partial \sigma^k(t_1+t_2) \bar{\sigma}^l(t_1)\Big>_L
\nn
&=&
\left(\frac{\pi}{L}\right)^{h+1} C_{-{\bm p}}^{l,k}
\frac{(2\pi)^2R_1R_2}{|\cos\pi\lambda|}\Bigg[
\left(\lambda-\frac{1}{2}\right)\alpha'\epsilon^{\mu\nu}p_\nu\,h\,
\frac{\sin^{h-1} \theta_{t_2}}{\sin^h \theta_{t_1+\frac{1}{2}}
\sin^{h+1} \theta_{t_3+\frac{1}{2}}}\times
\nn
&&~~~~~~~~~~\times\left(\cos \theta_{t_2}\sin \theta_{t_3+\frac{1}{2}}
+\sin\theta_{t_2}\cos \theta_{t_3+\frac{1}{2}}\right)
\nn
&&-\frac{i}{2}\alpha^{\prime}p^\mu
\frac{\sin^{h-2} \theta_{t_2}}{\sin^h \theta_{t_1+\frac{1}{2}}
\sin^{h+1} \theta_{t_3+\frac{1}{2}}}
\Big\{
-(h-1) \cos\theta_{t_2}\cos\theta_{t_1+\frac{1}{2}}
   \sin^2\theta_{t_3+\frac{1}{2}}
\nn
&&
+(h-1) \cos\theta_{t_2}\sin\theta_{t_1+\frac{1}{2}}
   \sin\theta_{t_3+\frac{1}{2}}\cos\theta_{t_3+\frac{1}{2}}
\nn
&&
-(h-1) \sin\theta_{t_2}\cos\theta_{t_1+\frac{1}{2}}
   \sin\theta_{t_3+\frac{1}{2}}\cos\theta_{t_3+\frac{1}{2}}
\nn
&&
+h \sin\theta_{t_2}\sin \theta_{t_1+\frac{1}{2}}
    \cos^2\theta_{t_3+\frac{1}{2}}
+ \sin\theta_{t_2}\sin\theta_{t_1+\frac{1}{2}}
   \sin^2\theta_{t_3+\frac{1}{2}}
\Big\}
\Bigg],
\end{eqnarray}
and
\begin{eqnarray}
&&
\Big<\partial X^\mu e^{-i{\bm p}\cdot {\bm X}}
\left(L-\frac{1}{2}\right)
\sigma^k(t_1+t_2) \partial \bar{\sigma}^l(t_1)\Big>_L
\nn
&=&
\left(\frac{\pi}{L}\right)^{h+1} C_{-{\bm p}}^{l,k}
\frac{(2\pi)^2R_1R_2}{|\cos\pi\lambda|}\Bigg[
-\left(\lambda-\frac{1}{2}\right)\alpha'\epsilon^{\mu\nu}p_\nu\,h\,
\frac{\sin^{h-1} \theta_{t_2}}{\sin^{h+1} \theta_{t_1+\frac{1}{2}}
\sin^h \theta_{t_3+\frac{1}{2}}}\times
\nn
&&~~~~~~~~~~\times\left(\cos \theta_{t_2}\sin \theta_{t_1+\frac{1}{2}}
+\sin\theta_{t_2}\cos \theta_{t_1+\frac{1}{2}}\right)
\nn
&&-\frac{i}{2}\alpha^{\prime}p^\mu
\frac{\sin^{h-2} \theta_{t_2}}{\sin^{h+1} \theta_{t_1+\frac{1}{2}}
\sin^h \theta_{t_3+\frac{1}{2}}}
\Big\{
-(h-1)\cos\theta_{t_2}\cos\theta_{t_3+\frac{1}{2}}
       \sin^2\theta_{t_1+\frac{1}{2}}
\nn
&&
+(h-1) \cos\theta_{t_2}\sin\theta_{t_3+\frac{1}{2}}
   \sin\theta_{t_1+\frac{1}{2}}\cos\theta_{t_1+\frac{1}{2}}
\nn
&&
-(h-1) \sin\theta_{t_2}\cos\theta_{t_3+\frac{1}{2}}
   \sin\theta_{t_1+\frac{1}{2}}\cos\theta_{t_1+\frac{1}{2}}
\nn
&&
+h \sin\theta_{t_2}\sin\theta_{t_3+\frac{1}{2}}
   \cos^2\theta_{t_1+\frac{1}{2}}
+ \sin\theta_{t_2}\sin\theta_{t_3+\frac{1}{2}}
   \sin^2\theta_{t_1+\frac{1}{2}}
\Big\}
\Bigg],
\end{eqnarray}
where $\theta_{s}$ is defined by $\theta_{s}=\pi s/L$. The ghost
correlation functions which appear in (\ref{profile_phi1}) are
given as 
\begin{eqnarray}
 \Big<c\partial c\left(L-\frac{1}{2}\right)
c(t_1+t_2) \Big>_L
&=& -\left(\frac{L}{\pi}\right)^2\sin^2\theta_{t_3+\frac{1}{2}},
\\
\Big<c\partial c\left(L-\frac{1}{2}\right)
c(t_1) \Big>_L
&=& -\left(\frac{L}{\pi}\right)^2\sin^2\theta_{t_1+\frac{1}{2}}.
\end{eqnarray}
Combining these results, we find that the integrand in (\ref{profile_phi1})
turns out to be 
\begin{eqnarray}
&& \Big<\partial X^\mu e^{-i{\bm p}\cdot {\bm X}}
\left(L-\frac{1}{2}\right)
\partial \sigma^k(t_1+t_2) \bar{\sigma}^l(t_1)\Big>_L
\Big<c\partial c\left(L-\frac{1}{2}\right)
c(t_1+t_2) \Big>_L
\nn
&&~~~~~-\Big<\partial X^\mu e^{-i{\bm p}\cdot {\bm X}}
\left(L-\frac{1}{2}\right)
\sigma^k(t_1+t_2) \partial \bar{\sigma}^l(t_1)\Big>_L
\Big<c\partial c\left(L-\frac{1}{2}\right)
c(t_1) \Big>_L
\nn
&=&
\left(\frac{\pi}{L}\right)^{h-1} C_{-{\bm p}}^{l,k}
\frac{(2\pi)^2R_1R_2}{|\cos\pi\lambda|}\Bigg[
-\left(\lambda-\frac{1}{2}\right)\alpha'\epsilon^{\mu\nu}p_\nu\,h\,
\frac{\sin^{h-1} \theta_{t_2}}{\sin^h \theta_{t_1+\frac{1}{2}}
\sin^h \theta_{t_3+\frac{1}{2}}}\times
\nn
&&~~~~~~~~~~\times\Big\{\cos \theta_{t_2}\Big(
\sin^2 \theta_{t_1+\frac{1}{2}}+\sin^2 \theta_{t_3+\frac{1}{2}}\Big)
+\frac{1}{2}\sin\theta_{t_2}
\Big(\sin 2\theta_{t_1+\frac{1}{2}}+\sin 2\theta_{t_3+\frac{1}{2}}
\Big)\Big\}
\nn
&&+\frac{i}{2}\alpha^{\prime}p^\mu
\frac{\sin^{h-2} \theta_{t_2}}{\sin^h \theta_{t_1+\frac{1}{2}}
\sin^h \theta_{t_3+\frac{1}{2}}}
\Big\{
(h-1)\cos\theta_{t_2}
      \Big(
 \sin^3\theta_{t_1+\frac{1}{2}}\cos\theta_{t_3+\frac{1}{2}}
- \sin^3\theta_{t_3+\frac{1}{2}}\cos\theta_{t_1+\frac{1}{2}}
\Big)
\nn
&&
-(h-1) \cos\theta_{t_2}
   \sin\theta_{t_1+\frac{1}{2}}\sin\theta_{t_3+\frac{1}{2}}
\Big(
\sin\theta_{t_1+\frac{1}{2}}\cos\theta_{t_1+\frac{1}{2}}
-\sin\theta_{t_3+\frac{1}{2}}\cos\theta_{t_3+\frac{1}{2}}
\Big)
\nn
&&
+(h-1) \sin\theta_{t_2}\cos\theta_{t_1+\frac{1}{2}}
   \cos\theta_{t_3+\frac{1}{2}}\Big(
   \sin^2\theta_{t_1+\frac{1}{2}}
  -   \sin^2\theta_{t_3+\frac{1}{2}}\Big)
\nn
&&-h \sin\theta_{t_2}\sin \theta_{t_1+\frac{1}{2}}
 \sin \theta_{t_3+\frac{1}{2}}\Big(
    \cos^2\theta_{t_1+\frac{1}{2}}
   -    \cos^2\theta_{t_3+\frac{1}{2}}\Big)
\nn
&&
- \sin\theta_{t_2}
   \sin\theta_{t_1+\frac{1}{2}}\sin\theta_{t_3+\frac{1}{2}}
\Big(\sin^2\theta_{t_1+\frac{1}{2}}
-\sin^2\theta_{t_3+\frac{1}{2}}\Big)
\Big\}
\Bigg].
\label{correlfunc_phi1}
\end{eqnarray}
The second term in the bracket on the right hand side proportional
to $p^{\mu}$ is antisymmetric under the interchange of $t_{1}$ and
$t_{3}$. Since the integration measure in (\ref{profile_phi1}) is
symmetric with respect to it, the second term does not contribute
to the vector profile.

Next let us turn to (\ref{profile_phi2}). The correlation function
which appears in the integrand in (\ref{profile_phi2}) is given as
\begin{eqnarray}
&&
\left(\frac{\pi}{L}\right)^{h-1} (h-1)\,C_{-{\bm p}}^{l,k}
\frac{(2\pi)^2R_1 R_2}{|\cos\pi\lambda|}\frac{\sin^{h-2}\theta_{t_2}}{
\sin^h \theta_{t_1+\frac{1}{2}}\sin^h \theta_{t_3+\frac{1}{2}}}\times
\nn
&&
\hspace{-.8cm}
\times\frac{1}{L}\Bigg[
\frac{\pi}{L}\sin\theta_{t_2}\Big\{
(1+2t_1)\cos\theta_{t_1+\frac{1}{2}}
  \sin\theta_{t_3+\frac{1}{2}}
-(1+2t_3)
\cos\theta_{t_3+\frac{1}{2}}\sin\theta_{t_1+\frac{1}{2}}
\Big\}
\nn
&&
+\frac{2\pi(t_1-t_3)}{L}\cos\theta_{t_2}\sin\theta_{t_1+\frac{1}{2}}
    \sin\theta_{t_3+\frac{1}{2}}
\nn
&&
-2\sin\theta_{t_2} \sin\theta_{t_1+\frac{1}{2}}
  \sin\theta_{t_3+\frac{1}{2}}
\Big(
\cos^2\theta_{t_1+\frac{1}{2}}
-\cos^2\theta_{t_3+\frac{1}{2}}
\Big)
\nn
&&
-2\cos\theta_{t_2}
\Big(
\cos\theta_{t_1+\frac{1}{2}}
  \sin^2\theta_{t_1+\frac{1}{2}}
    \sin\theta_{t_3+\frac{1}{2}}
-\cos\theta_{t_3+\frac{1}{2}}
  \sin^2\theta_{t_3+\frac{1}{2}}
\sin\theta_{t_1+\frac{1}{2}}
 \Big)\,\Bigg].
\label{correlfunc_phi2}
\end{eqnarray}
Since (\ref{correlfunc_phi2}) is antisymmetric under the interchange
of $t_{1}$ and $t_{3}$, we find that the right hand side of (\ref{profile_phi2})
vanishes.


\end{document}